\documentclass{aa}
\usepackage{graphicx}
\def\approxgt{\mathrel{\hbox{\rlap{\lower.55ex \hbox {$\sim$}}
        \kern-.3em \raise.4ex \hbox{$>$}}}}
\def\approxlt{\mathrel{\hbox{\rlap{\lower.55ex \hbox {$\sim$}}
        \kern-.3em \raise.4ex \hbox{$<$}}}}
\begin{document}
   \title{The hard X-ray view of the low-luminosity blazar \\ in
the radio galaxy NGC~6251}

   \author{M. Guainazzi,
          \inst{1}
          P.Grandi,
	  \inst{2}
	  A.Comastri,
	  \inst{3}
	  \and
	  G.Matt
          \inst{4}
          }

   \offprints{M.Guainazzi}

   \institute{XMM-Newton Science Operation Center, VILSPA, ESA, Apartado
              50727,E-28080 Madrid, Spain \\
              \email{mguainaz@xmm.vilspa.esa.es}
              \and
              Istituto di Astrofisica Spaziale e Fisica Cosmica (IASF-C.N.R.),
		Sezione di Bologna, Via Gobetti 101, I-40129 Bologna, Italy
	      \and
	      I.N.A.F., Osservatorio Astronomico di Bologna, via Ranzani 1, I-40127 Bologna, Italy 
	      \and
              Dipartimento di Fisica, Universit\`a degli Studi `Roma Tre', Via della Vasca Navale 84, I-00146 Roma, Italy	
              }

   \date{Received ; accepted }

   \abstract{We present results from a BeppoSAX (July 2001)
observation of the
FR~I radio galaxy NGC~6251, together with a re-analysis of 
archival ASCA
(October 1994) and
{\it Chandra} (September 2000)
data. The weak detection above 10~keV
and the lack of iron fluorescent K$_{\alpha}$ emission lines 
in the BeppoSAX spectrum rule
out that the bulk of the X-ray emission is due to an
obscured Seyfert nucleus. The study of the multiwavelength spectral
energy distribution suggests instead that X-rays probably originate
as inverse-Compton of synchrotron seed photons in a relativistic
jet, indicating that NGC~6251 hosts a
low radio luminosity ($L_{{\rm 5 \ GHz}} \sim 10^{40}$~erg~s$^{-1}$)
blazar. The BeppoSAX spectrum is flatter than
in the earlier ASCA observation. This might be due to the emergence of
a different spectral component during phases of lower
X-ray flux.
In this context, we discuss some possible explanations for the intense
and mildly-ionized fluorescent iron line measured by ASCA. 
            }

\authorrunning{Guainazzi et al.}

\titlerunning{The hard X-ray view of NGC~6251}

   \maketitle
%

\section{Introduction}

NGC~6251 is an E2 nearby ($z=0.02488$; if
$H_{\rm 0} = 70$~km~s$^{-1}$~Mpc$^{-1}$, as
assumed in this paper, 1$\arcsec$ corresponds
to about 500~pc) radio galaxy with a Faranoff-Riley~I
morphology, known to host
a giant radio jet (\cite{wagget77}), a Seyfert~2 nucleus
(\cite{werner00}), and almost edge-on dust lanes (\cite{nieto83}).
The nucleus is likely to contain a supermassive black hole
with mass $M \sim 4$--$8 \times 10^8 M_{{\rm \odot}}$
(\cite{ferrarese99}), as
suggested by the presence of nuclear gas and a dust disc
on scales of a few hundreds parsecs.
NGC~6251 belongs to the outskirts of the cluster Zw 1609.0+8212
(\cite{young79}), whose influence on the properties of the
galaxy should be, however, marginal (\cite{prestage88}).

The NGC~6251 jet is one of the most spectacular
radio objects of the whole sky.
It is a 4$\arcmin$.5 long, highly collimated
(opening angle 7.4$^{\circ}$) structure
(\cite{perley84}). Regions of
enhanced radio emission along the jet
were labeled  by Birkinshaw \& Worrall (1993) in the 330~MHz
radio map
as A (10--40$\arcsec$ from the
nucleus), B (40--126$\arcsec$;
actually structured in smaller sub-structures)
and D (178-264$\arcsec$, where the
jet bends towards the North;
it is as well highly structured).
Between B and D one finds a region of low radio
brightness (C).
None of these region was observed to be a significant
source of X-ray emission by ROSAT
(\cite{birkinshaw93}). A sub-pc
counter-jet was discovered
only recently with Very Long Baseline Interferomer
observations (\cite{sudou01}). U-band
{\it Hubble Space Telescope}
images unveiled a region of extended
emission, lying nearly perpendicular to the radio
jet axis and the dust ring
(\cite{crane97}), which is likely to
originate from scattering of a nuclear continuum
source.

Discovered in the X-rays by the {\it Einstein}
Imaging Proportional Counter (\cite{jones86}),
NGC~6251
showed in the ROSAT PSPC observation (\cite{birkinshaw93})
an unresolved core ($FWHM < 4\arcsec$), with a possible
extended halo on scales $\approxlt 100$~kpc (\cite{mack97}).
The first
observation of NGC~6251 covering the intermediate X-rays
(i.e.: 2--10~keV) was performed by ASCA.
The presence of
a bright K$_{\alpha}$ fluorescence iron from
ionized iron (centroid energy, $E_c \simeq 6.68$~keV;
Equivalent Width, $EW \simeq 600$~eV;
\cite{turner97}; \cite{sambruna99}),
and the fact that the continuum  could be best fit
with a combination of a standard AGN power-law
component (photon index, $\Gamma$, of 2.11)
plus a thermal soft excess, suggested that ionized
gas may significantly contribute in this energy band.
Electron scattering of an otherwise
invisible nuclear continuum was an interesting
possibility, in light of the HST discovery of extended
ionized gas, which could potentially act as
a nuclear mirror (``warm mirror" hereinafter).
Recently, an association has been proposed between
NGC~6251 and the EGRET source 1EGJ1621+8203
(\cite{mukherjee02}). If this is confirmed,
NGC~6251 would be one of the three radio galaxies -
alongside with Cen~A (\cite{sreekumar99})
and 3EG~J1735-1500 (\cite{combi03}) - detected in high-energy
$\gamma$-rays. No detection by the {\it Extreme
Ultraviolet Explorer} is reported in the literature
(\cite{marshall95}).

The deep BeppoSAX observation, described in this paper,
aimed at verifying the
interpretation scenario emerging from the ASCA outcomes.
Thanks to its
unsurpassed sensitivity in
hard X-rays (i.e.: $>$10~keV; \cite{boella97a}),
BeppoSAX would be easily able to detect a transmitted
nuclear component piercing through
a Compton-thick absorber (\cite{matt00}).
The results of this observation are presented
in this paper, together with
an analysis of archival ASCA and {\it Chandra}
data of the same source. The log of the observations
discussed in this paper is presented in Table~\ref{tab4}.
\begin{table*}
\begin{center}
\begin{tabular}{lccc} \hline \hline
Satellite & Start time & Exposure time & Count rate \\
& & (ks) & ($10^{-2}$~s$^{-1}$) \\ \hline
ASCA & 28-Oct-1994 & 36$^a$/44$^b$ & $7.2 \pm 0.2$$^a$/$3.63 \pm 0.13$$^b$\\
{\it Chandra}/ACIS-I & 11-Sep-2000 & 25 & $8.1 \pm 0.2$ \\
BeppoSAX & 19-Jul-2001 & 28$^c$/80$^d$/72$^e$ & $2.78 \pm 0.12$$^c$/$5.55 \pm 0.09$$^d$/$13 \pm 4$$^e$ \\ \hline \hline
\end{tabular}
\end{center}

\noindent
$^a$SIS0,
\noindent
$^b$GIS2,
\noindent
$^c$LECS,
\noindent
$^d$MECS,
\noindent
$^e$PDS.

\caption{Log of the observations presented in this paper.}
\label{tab4}
\end{table*}

While this paper was in the process of being accepted,
we became aware of a recently accepted paper by Chiaberge et
al. (2003), who discuss the same X-ray observations,
together with literature and HST data. Their X-ray results
are largely coincident with ours.

\section{BeppoSAX results}

\subsection{Data reduction}

BeppoSAX data were reduced according to standard
procedures as in, e.g., Guainazzi et al. (1999).
Scientific products for the imaging Low
Energy Concentrator Spectrometer (LECS, \cite{parmar97},
0.5--4~keV) and Medium Energy Concentrator Spectrometer
(MECS, \cite{boella97b}, 1.8--10.5~keV) were
extracted from circular regions of 8$\arcmin$ and 4$\arcmin$,
respectively. Background spectra were extracted
from blank sky field event lists provided by the
BeppoSAX Science Data Center, and appropriate for the
date of the observation. Products for the
Phoswitch Detector System (PDS, \cite{frontera97},
13--200~keV) were extracted by plain subtraction of those
corresponding to intervals when the instrument
was observing NGC~6251, and a region $\pm$3.5$^{\circ}$ aside,
according to the standard 96~s cycle.

All the spectra employed in this paper have been
rebinned in order to oversample the intrinsic energy
resolution of the instruments by a factor not larger than
3, and to have a number of counts in each spectral channel
higher than 30, in order to ensure the applicability
of the $\chi^2$ test. In this paper:
energies are quoted in the
source rest frame; uncertainties
on the spectral parameters 
are quoted at the 90\% confidence level for one interesting parameter;
uncertainties on the count rates are at 1$\sigma$ level,
unless otherwise specified. 

\subsection{Spectral results}

\begin{figure}
   \centering
   \includegraphics[angle=-90,width=8cm]{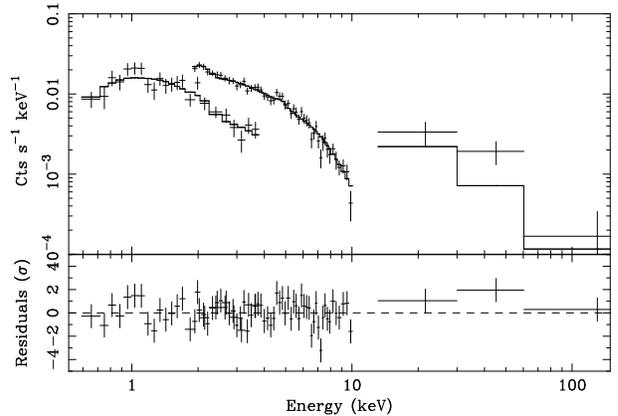}
      \caption{BeppoSAX spectrum ({\it upper panel})
	      and residuals in units of standard
	      deviations ({\it lower panel}), when
	      the a photoelectrically absorbed
	      power-law model is applied.
              }
         \label{fig1}
   \end{figure}

In Fig.~\ref{fig1} the results of the fit of the NGC~6251
BeppoSAX spectrum with a simple absorbed
power-law is shown. The
fit is statistically acceptable ($\chi^2 = 96.7/92$ degrees
of freedom, dof). The best-fit
parameters and results are reported in Table~\ref{tab1}.
\begin{table*}
\begin{center}
\begin{tabular}{lcccccccc} \hline \hline
Model & $N_H$ & $\Gamma$ & $kT_1$/$\Gamma_{{\rm soft}}$ & $kT_2$/$E_{{\rm break}}$ & $E_c$ & $I_c$ & $EW$ & $\chi^2/$~dof \\
Mission & ($10^{21}$~cm$^{-2}$) & & (keV)/ & (keV) & (keV) & ($10^{-6}$~ph~cm$^{-2}$~s$^{-1}$) & (eV) & \\ \hline
\multicolumn{7}{l}{{\it BeppoSAX}} \\
PL & $0.8 \pm^{0.6}_{0.3}$ & $1.79 \pm^{0.06}_{0.07}$ & ... & ... & ... & ... & ... & 96.7/92 \\
PL+TH & $1.0 \pm^{0.12}_{0.05}$ & $1.70 \pm^{0.12}_{0.16}$ & $1.4 \pm^{0.9}_{0.7}$ & ... & 6.4$^{\dag}$ & $<8.0$ & $< 160$ & 92.3/90 \\
 & &  & & ... & 6.59$^{\dag}$ & $<4.5$ & $<80$ &  \\
 & &  & & ... & 6.7$^{\dag}$ & $<3.0$ & $<60$ &  \\
 & &  & & ... & 6.96$^{\dag}$ & $<2.1$ & $<50$ &  \\ \hline
\multicolumn{7}{l}{{\it ASCA}} \\ 
PL & $0.9 \pm^{0.4}_{0.3}$ & $2.06 \pm^{0.08}_{0.07}$ & ... & ... & ... & ... & ... &  259.7/225 \\
PL+TH & $1.6 \pm^{0.7}_{0.8}$ & $2.00 \pm^{0.08}_{0.07}$ & $0.85 \pm^{0.34}_{0.15}$ & ... & ... & ... & ... &  240.6/223 \\
PL+2$\times$TH & $1.6 \pm^{2.1}_{0.8}$ & $2.3 \pm 0.4$ & $0.8 \pm^{0.3}_{0.2}$ & $6 \pm^6_2$ & ... & ... & ... &  231.8/221 \\
PL+TH+GA & $1.7 \pm^{0.5}_{0.4}$ & $2.05 \pm^{0.11}_{0.10}$ & $0.84 \pm^{0.27}_{0.18}$ & & $6.59 \pm 0.16$ & $8 \pm 4$ &  $500 \pm 200$ & 228.1/221 \\
BKPL+GA & $4.0 \pm^{1.8}_{1.5}$ & $2.28 \pm^{0.17}_{0.14}$ & $3.7 \pm^{1.2}_{0.9}$ & $1.38 \pm^{0.15}_{0.09}$ & $6.62 \pm^{0.12}_{0.18}$ & $10 \pm4 $ & $800 \pm 300$ & 245.0/226 \\ \hline \hline
\end{tabular} 
\end{center}

\noindent
$^{\dag}$Gaussian emission profiles added to the best-fit continuum. Centroid energies $E_c$ are fixed. Upper limits are at the 90\% confidence level

\caption{Best-fit parameters and results for the analysis of the
nuclear spectra of NGC~6251. Model legenda: PL~=~power-law; TH~=~thermal
component ({\tt mekal} implementation in {\sc Xspec})
with solar abundances; GA~=~Gaussian
emission line; BKPL~=~broken power-law.}
\label{tab1}
\end{table*}
The PDS data points lay actually above the
best-fit model, which is mainly determined by the statistics
of the imaging instruments. The excess is, however, not highly
significant. The ratio between the PDS flux and the
extrapolation of the best fit model in the LECS/MECS energy
band is $1.4 \pm 0.6$, therefore only marginally exceeding
the typical values of the cross-normalization
factors
between the PDS and the MECS (0.80--0.85; \cite{fiore98}).
Not surprisingly,
several models provide comparably good descriptions of this
hard excess, but none of them is required from the
statistical point of view. A broken power-law, for instance,
yields a $\Delta \chi^2=3.4$ for a reduction
of the number of degrees of freedom ($\Delta \nu$) by 2
(this quantity will be indicated as $\Delta \chi^2/\Delta \nu$
hereinafter),
corresponding to a confidence level $\simeq$83.0\%

The BeppoSAX large source spectrum extraction region
encompasses the putative ROSAT extended emission.
We therefore tried to add a
soft X-ray component, modeled by a collisionally ionized plasma
(we used the {\tt mekal} implementation in {\sc
Xspec v11.0} throughout this paper; \cite{mewe85},
\cite{liedahl95}, \cite{arnaud92}). Again,
the improvement in the quality of the fit is not
significant ($\Delta \chi^2 / \Delta \nu = 4.4/2$). Consequently,
the temperature is rather poorly determined
($kT = 1.4 \pm^{0.9}_{0.7}$). The 0.5--2.4~keV
flux of this component would be
$(1.3 \pm 0.7) \times 10^{-13}$~erg~cm$^{-2}$~s$^{-1}$, corresponding
to about 15\% of the non-thermal component in the same
energy band. This model requires an absorbing column
density slightly in excess with respect to the Galactic
contribution along the line-of-sight to NGC~6251
($5.7 \times 10^{20}$~cm$^{-2}$, \cite{murphy96}).

No systematic residual feature is present
at the energies, where K$_{\alpha}$ fluorescence transitions
of iron are expected. Negligible improvements of the $\chi^2$
are yielded by the addition of a narrow
(i.e., intrinsic width, $\sigma = 0$) Gaussian profile to the
power-law model. 90\% upper limits on the intensity
of a neutral (6.4~keV) or He-like (6.7~keV) iron lines
are 8.5 and $3.0 \times 10^{-6}$~photons~cm$^{-2}$~s$^{-1}$,
respectively, corresponding to EWs of
160 and 60~eV, respectively. These upper limits are largely
inconsistent with the ASCA detection (\cite{turner97};
cf. also Sect.~3.1 later). Simulations show that a 600~eV
EW line would have been detected at the 6-7$\sigma$ confidence
level in the MECS spectrum.

The 0.5--10~keV flux during the BeppoSAX observation was
$7.5 \times 10^{-12}$~erg~cm$^{-2}$~s$^{-1}$, corresponding
to un unabsorbed rest-frame luminosity of
$1.13 \times 10^{43}$~erg~s$^{-1}$
in the same energy band.

\section{Comparison with previous X-ray observations}

\subsection{ASCA}

We have retrieved the data of the ASCA observation of
NGC~6251 from the public archive as
screened event lists. Spectra were extracted from regions
of radii 4$\arcmin$, 3$\arcmin$.1 and 3$\arcmin$.75
in the SIS0, SIS1 (grade 0, 2, 3, 4)
and GIS detectors, respectively
(they therefore encompass as well the ROSAT
extended emission
region).
Background spectra were extracted from regions in
the field of view of the detectors, free from
contaminating sources. Response matrices
were generated with the packages included
in the {\sc Lheasoft} v5.0
software.
The spectral analysis was
performed in the 0.5--10~keV and
0.7--10~keV energy bands, for the SIS and GIS instruments,
respectively. The results of our analysis substantially
coincide with those presented by Turner et al. (1997)
and Sambruna et al. (1999), and
we summarize them in this paper for
the sake of clarity only.

\begin{figure}
   \centering
   \includegraphics[angle=-90,width=8cm]{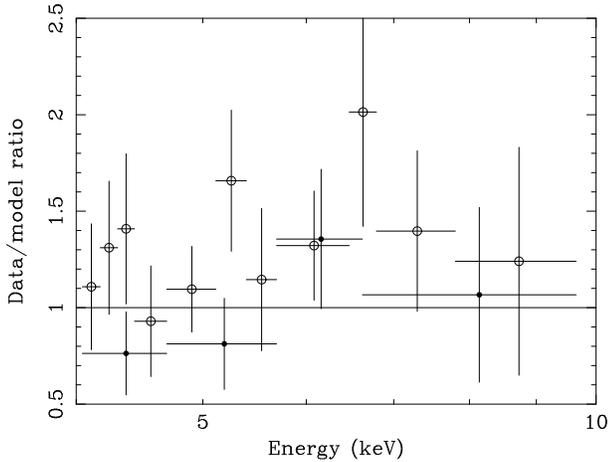}
      \caption{GIS2 ({\it filled dots}) and GIS3
		({\it open circles}) residuals against
	        a power-law plus optically thin thermal
		plasma continuum in the 4--10~keV bands.
              }
         \label{fig2}
   \end{figure}

A simple power-law is a fairly good representation of the
ASCA spectra ($\chi^2 = 259.7/225$~dof). However,
the addition of a {\tt mekal} component improves
significantly
the quality of the fit ($\Delta \chi^2/\Delta \nu = 19.1 / 2$,
corresponding to an F-test confidence level $\simeq$$99.97\%$).
The addition of a
multitemperature blackbody
(model {\tt diskbb} in {\sc Xspec}) yields, on the contrary,
a negligible
improvement to the quality of the fit. Modeling
the continuum in terms of a broken power-law yields
a comparatively worse fit as well.
A systematic excess around the energy where
fluorescent K$_{\alpha}$ transitions from iron are
expected is observed (see Fig.~\ref{fig2}).
It can be cured with an
unresolved Gaussian profile ($\Delta \chi^2/ \Delta \nu = 12.5/2$,
corresponding to a confidence level $\simeq$$99.7\%$),
or with a hotter ($kT \simeq 6$~keV) thermal
component ($\Delta \chi^2 / \Delta \nu = 8.8/2$,
corresponding to a confidence level $\simeq$$98.6\%$).

In the best-fit model (cf. Table~\ref{tab1}),
the power-law index 
($\Gamma \simeq 2.06$) is
significantly steeper than
observed by BeppoSAX. The temperature
of the colder thermal component is $kT \simeq 0.8$~keV,
and 
its 0.5--2.4~keV flux
[$(1.6 \pm 0.4) \times 10^{-13}$~erg~cm$^{-2}$~s$^{-1}$]
is in good agreement with that of the corresponding
component in the BeppoSAX spectrum.
The iron line centroid energy is consistent with
K$_{\alpha}$ fluorescence from mildly ionized to
He-like iron. Its $EW$
is $500 \pm 200$~eV, significantly larger than
the upper limits measured by BeppoSAX.
The difference is most likely due to a change
in the underlying continuum, as the
intensity of the ASCA line 
[$ (8 \pm 4) \times 10^{-6}$~photons~cm$^{-2}$~s$^{-1}$] is
marginally consistent with the BeppoSAX upper limits.
However, an intrinsic variability of the iron line cannot
be ruled out.

The observed 0.5--10~keV flux during the ASCA
observation was
$2.7 \times 10^{-12}$~erg~cm$^{-2}$~s$^{-1}$,
corresponding to an unabsorbed total luminosity of
$4.4 \times 10^{42}$~erg~s$^{-1}$
in the same energy band.

\subsection{{\it Chandra}}

Data of an ACIS-I NGC~6251 {\it Chandra}
observation
were retrieved from the public archive.
A bright source
[count rate $(8.1 \pm 0.2) \times 10^{-2}$~s$^{-1}$]
is detected with coordinates:
$\alpha_{{\rm 2000}}=16^h32^m31^s.8$;
$\delta_{{\rm 2000}}=+82^h32^m16^s$, i.e. 0.6$\arcsec$
distant from the optical nucleus of NGC~6251,
well within the accuracy of the {\it Chandra} attitude
reconstruction.
At this count rate level, a {\it Chandra} source is likely
to be substantially affected by pile-up, given the
instrumental mode employed (Time Exposure Mode
with a 3.2~s frame time). We have tried to fit the
0.5--8~keV
spectrum extracted from the innermost 5$\arcsec$
with a photoelectrically absorbed power-law, corrected
for pile-up according to the {\sc Xspec} implementation
of J.Davis' algorithm (model {\tt pileup} in
{\sc Xspec}; \cite{arnaud02}). The fit is
acceptable ($\chi^2 = 57.1/78$~dof), and yields
best-fit parameters which are consistent
with those measured during the BeppoSAX observation,
save a 60\% lower flux: $N_H = (1.6 \pm 0.5) \times
10^{21}$~cm$^{-2}$; $\Gamma = 1.76 \pm 0.16$;
0.5--10~keV flux of
$4.5 \times 10^{-12}$~erg~cm$^{-2}$~s$^{-1}$.

Thanks to the unprecedented spatial resolution of the
ACIS-S, the issue of the spatial extension can be
better addressed.
No evidence for extended emission
along the arcminutes scale jet is detected,
as the ROSAT observation had already shown in the
soft X-rays only (\cite{birkinshaw93}).
The upper limits on the X-ray
fluxes
of the regions A to D are
reported in Table~\ref{tab2}.
\begin{table}
\begin{center}
\begin{tabular}{lccccc} \hline \hline
Region & $d$ & Area & $CR$ & $F$ & $S_{\nu}$ \\
& (arcsec) & (arcsec$^2$) & ($10^{-4}$) & $^a$ & (nJy) \\ \hline
A & 10-40 & 46.8 & 2.8 & 2.3 & 0.34 \\
B & 40-126 & 140.6 & 2.2 & 1.8 & 0.26 \\
C & 126-178 & 78.1 & 2.9 &  2.4 & 0.36 \\
D & 178-264 & 124.9 & 1.9 & 1.6 & 0.24 \\ \hline \hline
\end{tabular}
\end{center}

\noindent
$^a$in the 0.5--10~keV band, in units of $10^{-15}$~erg~cm$^{-2}$~s$^{-1}$

\caption{90\% upper limits on the ACIS-I count rates ($CR$),
0.5--10~fluxes ($F$), and 1~keV flux density ($S_{\nu}$)
across the NGC~6251 jet regions
(following their definition in Birkinshaw \& Worrall 1993).
The distance $d$ is defined from the inner-outer border of
each region to the core.
Fluxes are calculated assuming a power-law spectrum, with
$\Gamma=2$ and photoelectric absorption
column density $N_H = 5.7 \times 10^{20}$~cm$^{-2}$.}
\label{tab2}
\end{table}
They are by a factor 5 to 10 tighter than those determined
by Birkinshaw \& Worrall (1993). However, the upper limit
on the knot ``D" flux density is inconsistent with
the detection ($S_{\nu} = 13 \pm 2$~nJy)
obtained by Mack et al. (1997) from
a reanalysis of the same ROSAT/PSPC observation
discussed by  Birkinshaw \& Worrall (1993).

In
the innermost 4$\arcmin$ around
the NGC~6251 core only two sources, alongside with
the nucleus itself, are detected at a signal-to-noise
ratio higher than 3. Their positions and count rates
are reported in Table~\ref{tab3}.
\begin{table}
\begin{center}
\begin{tabular}{lccc} \hline \hline
\# & $\alpha_{{\rm 2000}}$ & $\delta_{{\rm 2000}}$ & count rate \\
& & & ($10^{-4}$~s$^{-1}$) \\ \hline
1 & $16^h32^m31^s.8$ & $+82^h32^m16^s$ & $810 \pm 20$$^a$ \\
2 & $16^h33^m00^s.5$ & $+82^h31^m12^s$ & $4.8 \pm 1.8$ \\
3 & $16^h33^m21^s.3$ & $+82^h31^m56^s$ & $5.6 \pm 1.8$ \\ \hline \hline
\end{tabular}
\end{center}

\noindent
$^a$NGC~6251 nucleus

\caption{Sources detected in the ACIS-I observation of
the NGC~6251 field, within 4$\arcmin$ from the core.
}
\label{tab3}
\end{table}
None of them exhibits a clear association with any radio
structures.

\section{Discussion}

\subsection{The nature of the nuclear
X-ray emission in NGC~6251}

\begin{figure}
   \centering
   \includegraphics[width=8cm]{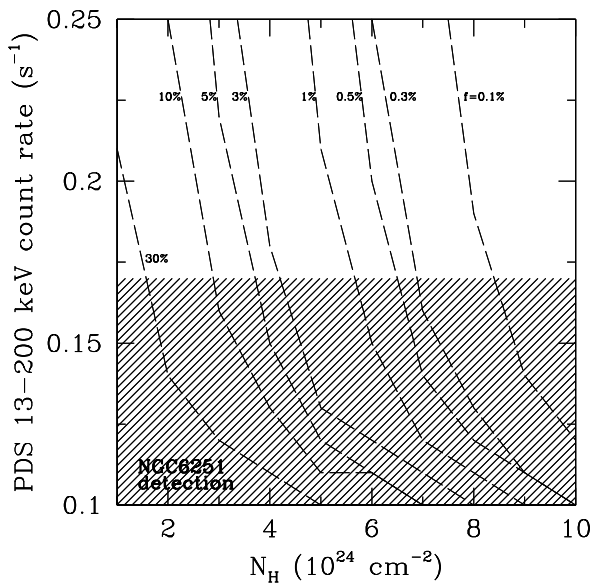}
      \caption{PDS count rates in the 13--200~keV
		energy band against the column density
		of a Compton-thick absorber covering the
		NGC~6251 nucleus, for different values
		of the scattering fraction $f$. The
		{\it shaded area} indicates the
		PDS detection yielded by the BeppoSAX
		observation of NGC~6251. $f$ is defined
		by the ratio between the normalizations
		of the transmitted and the warm scattered
		components, assuming an optically thin
		scatterer
              }
         \label{fig9}
   \end{figure}

The main goal of the BeppoSAX observation described in this
paper was to test whether the
strong ($EW \sim 600$~eV), ionized K$_{\alpha}$ fluorescent
iron line observed in the ASCA spectrum (\cite{turner97})
was due to a "warm mirror" reflection-dominated,
Compton-thick Seyfert~2 spectrum.
In the light of the BeppoSAX observation outcomes,
this possibility is rather unlikely.
The PDS detection ($0.13 \pm 0.04$~s$^{-1}$ in the
13--200~keV energy band) is statistically consistent
with the extrapolation of the 2--10~keV spectrum.
This constraints the column density of any Compton-thick absorber
to the nucleus to be $\approxgt 3 \times 10^{24}$
($6 \times 10^{24}$)~cm$^{-2}$ for a 10\% (1\%)
scattering fraction (see Fig.~\ref{fig9}).
Moreover, the
2--10~keV BeppoSAX spectrum lacks any trace of fluorescence
K$_{\alpha}$ ionized
iron lines, which are almost always observed
in reflection-dominated Seyfert~2 galaxies (\cite{turner97b},
\cite{matt00}).
The upper limits on their EWs are rather strict,
ranging between 60 and 80~eV for He- and H-like iron.
The standard AGN 2--10~keV
spectral index ($\Gamma \simeq 1.7$), and the lack of
K$_{\alpha}$ neutral iron line
(again, with a rather tight EW upper limit: 160~eV)
in the BeppoSAX spectrum rule out a significant
contribution by Compton-reflection from the inner
side of the molecular torus
(\cite{krolik94}, \cite{ghisellini94}, \cite{matt96})
In principle, the line flux
could be suppressed by resonant trapping at the ionized
surface of an accretion disk (\cite{matt93}).
However, further pieces of evidence
rule out that
the X-ray spectrum is dominated by reflection of an otherwise
invisible Seyfert nucleus.

\begin{figure}
   \centering
   \includegraphics[width=8cm]{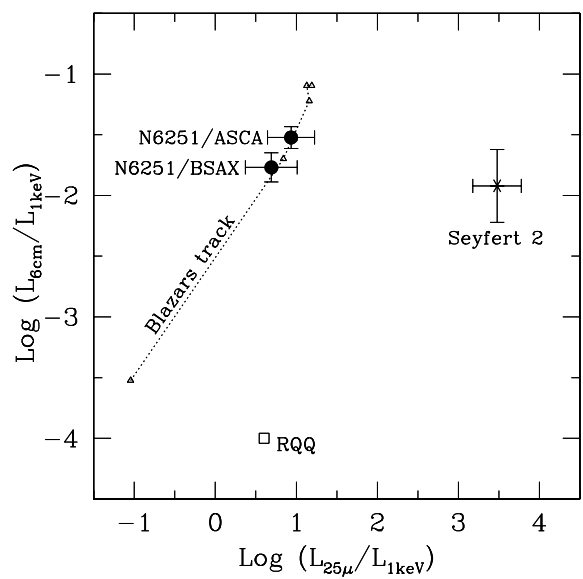}
      \caption{Radio (6~cm), X-ray (0.5--4.5~keV) and IR
		(25$\mu$) color-color diagram.
		The {\it dotted line} represents
		the blazar track according to the 
		blazar unified scheme of Fossati et al.
		(1998). The {\it
		filled circles} represent the data
		points corresponding to the NGC~6251
		ASCA and BeppoSAX observations. The
		{\it cross} corresponds to the
		Seyfert~2 sample of Mass-Hesse et al. (1995).
		The {\it empty square}
		represents the SED of radio-quiet
		quasars
		(\cite{elvis94})
              }
         \label{fig6}
   \end{figure}

Some hints come from the study of the multiwavelength
Spectral Energy Distribution (SED).
Recently, Fossati et al. (1998) have proposed a unified
scheme to explain the multivawelenght SED of blazars,
whose properties (peak frequency of
the synchrotron and inverse-Compton components, luminosity ratio
between them) are mainly governed by a
single parameter related to the overall luminosity.
This scheme can be applied to help identifying the
nature of the bulk of the X-ray emission from the
NGC~6251 nucleus. In Fig.~\ref{fig6} we show a
color-color diagram between the radio ($\nu = 6$~cm), soft
X-ray (0.5--4.5~keV), and infrared
($\lambda = 25 \mu$) luminosities. The dotted
line in this diagram represents the trace followed by blazars
according to the Fossati
scheme.
In the same plot the cross represents the
position of the Seyfert~2 galaxies of the Mass-Hesse et al. sample
(1995). The 6~cm radio luminosity of NGC~6251 is
$(1.0 \pm 0.1) \times 10^{40}$~erg~s$^{-1}$
(\cite{jones86}). The
data points corresponding to the ASCA and BeppoSAX
observations lay
intriguingly well on  the
blazar track. This supports the idea that
the bulk of the nuclear emission in
NGC~6251 is due to a jet. 

Chiaberge et al. (1999) discovered a clear correlation
between the optical ($F_O$) and the radio core luminosity
($F_R$) in a
sample of FRI galaxies extracted from the 3C catalogue
and observed with the HST WFC2. Assuming the NGC~6251
SED published by Ho (1999),
$\log (F_R/F_O) \simeq 3.4$,
in perfect agreement with the value derived from the
3C correlation ($3.7 \pm 0.4$; the uncertainties
represent the r.m.s. scattering of the data points
in the correlation), and largely
inconsistent with values typically measured
in radio-loud quasars ($\simeq$-1; \cite{elvis94}).
This supports a common origin
for the radio and optical emission as
synchrotron radiation. HST observations in the U-band
measured a rather high degree of polarization
(close to 50\%) in clumps close to the nucleus along
the radio axis. Crane \& Vernet (1997) suggest that
the UV emission of these clumps is due to scattering.
However, such an evidence is consistent as well
with the possibility
that the synchrotron-dominance extends well within the
UV range.

\begin{figure}
   \centering
   \includegraphics[width=8cm]{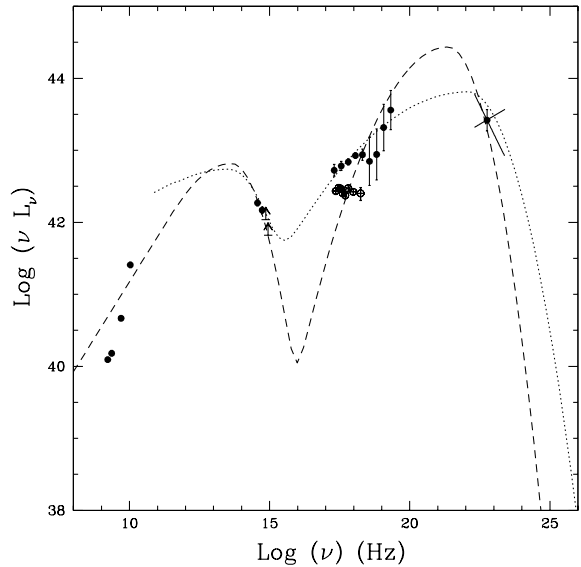}
      \caption{NGC~6251 SED.
		Data are from not-simultaneous observations
		compiled by Ho (1999), save the and
		X-ray data (BeppoSAX, {\it dots}; this paper), and
		the putative EGRET detection. The {\it
		dashed line} represents the best radio-$\gamma$
                fit SSC model; the {\it dotted line} represents
		the fit with an SSC model, forced to
		reproduce the BeppoSAX spectral shape. 
               The ASCA spectrum ({\it empty circles})
		is shown as well for reference
              }
         \label{fig10}
   \end{figure}

We tried to fit the overall NGC~6251 SED from radio to
$\gamma$-rays (X-rays represented by the BeppoSAX spectrum)
with an
homogeneous Synchrotron Self-Compton (SSC) model
(\cite{tavecchio98}).
This model assumes that synchrotron radiation is
produced by relativistic electrons
with density $n_e$
continuously injected
in a spherical region
of radius $r$ with a magnetic field $B$
and moving with bulk factor $\Gamma_{{\rm bulk}}$
at an angle $\theta$ with respect to the line of
sight (the Doppler factor is therefore
$\delta = [ \Gamma_{{\rm bulk}} (1- \beta \cos \theta)]^{-1}$).
These photons are subsequently upscattered by the same
electrons. Following Tavecchio et al. (1998), the
electron energy distribution is modeled with a
broken power-law with indices $n_1 <3$ and $n_2 > 3$
below and above a break energy $\gamma_b$, respectively.
The IRAS data points, although in principle available,
have not been included in the fit due to the unknown
contamination from the host galaxy.
The best fit from radio to $\gamma$ frequencies 
({\it dashed line} in Fig.~\ref{fig10})
fails to reproduce the BeppoSAX spectral shape and slightly deviates
below $10^{10}$ Hz.
If one forces the model to match the BeppoSAX data
point ({\it dotted line} in Fig.~\ref{fig10}),
the radio emission is more overproduced.
This may indicate that self-absorption
(which is not explicitly included in SSC models)
plays
an important role redwards the synchrotron peak.
Given the fact that the SED data points correspond
to non-simultaneous observations, one cannot push too far
the comparison between the model and the observation.
Nonetheless, the overall
properties of the best-fit SSC models are not
strongly dependent on the details of the fit. The
synchrotron peak is below $10^{14}$~Hz,
whereas the Compton peak remains constrained
between $10^{21.5}$ and $10^{22.5}$~Hz. 
As a reference
the best-fit parameters for the fits are in the range:
$B = 0.01$--0.15~G, $\delta = 2$--4, $n_1 = 1.75$--2.6,
$n_2 = 4.1$--5.4, $\gamma_b = 6.3 \times 10^3$--$1.6 \times 10^4$
$n_e=10^5$--10$^6$~cm$^{-3}$, and $r = 1$--8$\times10^{16}$~cm. 
The reader is referred to
Chiaberge et al. (2003) for a more detailed comparison
between the NGC~6251 SED and SSC models.

It is hard to interpret the spectral variability between
the ASCA ({\it empty circles} in Fig.~\ref{fig10})
and the BeppoSAX spectra in terms of pure SSC
model. In principle, the ASCA spectrum may represent the
trailing edge of the synchrotron component.
However, this would imply a shift of the synchrotron
peak by more than 4 orders of magnitude, which is rather
unlikely. Alternatively, the steeper ASCA spectrum may imply
that a different spectral component may be
emerging during phases of low X-ray flux. This
component may be completely outshined
during BeppoSAX-like, X-ray brighter states. We will
further discuss this possibility in Sect.4.2, in
connection with the strong fluorescent iron line
observed in the ASCA spectra only.

\begin{figure}
   \centering
   \includegraphics[width=8cm]{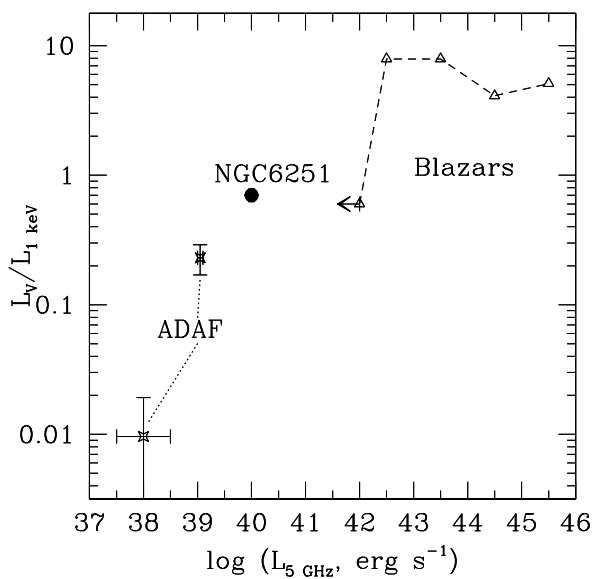}
      \caption{Radio between the 1~keV and the V-band
		flux density in: a) NGC~6251 ({\it
		filled circle}); b) the blazar
		radio luminosity classes in the
		unified scenario after Fossati
		et al. (1998); c) ADAF models
		applied to a sample of nearby
		elliptical galaxies (\cite{dimatteo00};
		details in text). In the last case,
		the error bars on the y-axis represent
		the r.m.s. of the sample values
		in the corresponding
		radio luminosity interval.
              }
         \label{fig8}
   \end{figure}

Recently, it has been suggested that the bulk of the
X-ray emission in radio galaxies may be due to
a hot inner accretion flow, following a line of thought 
suggested  more than 20 years
ago by Rees et al. (1982).
Ho (1999) estimates the bolometric luminosity of
NGC~6251 as $\sim$$10^{-4}$~L$_{{\rm Edd}}$. This may
indicate  that an Advection Dominated Accretion Flow
(ADAF; \cite{narayan95})
is responsible for the bulk of the emission in the
NGC~6251 core. In this scenario, X-rays are mainly
produced via bremsstrahlung by a distribution of
thermal electrons with typical temperatures
$kT \sim 100$~keV (\cite{dimatteo00}).
In Fig.~\ref{fig8} we compare
the ratio between the V and the 1~keV flux density
(BeppoSAX measurement)
in NGC6251 with the values expected by the blazar
unification scenario (\cite{fossati98}) and
by models of ADAF applied to the multiwavelength SED
of nearby elliptical galaxies suspected to host
low radiative efficiency accretion flows (\cite{dimatteo00};
we consider hereby models with wind, and outermost
hot accretion radius of 300 gravitational radii).
The NGC~6251 data point is in principle consistent
with both scenarios.
Applying a bremsstrahlung model to the BeppoSAX spectrum,
one gets indeed
a total X-ray luminosity $L_X/L_{{\rm Edd}} = (2 \pm 1) \times
10^{-3}$. However, the fit is significantly worse than
with a simple power-law ($\chi^2 = 110.0/92$~dof).
A composite bremsstrahlung and power-law fit (the latter
component taking into account possible Comptonization effects)
yields an implausibly low electron temperature ($kT \simeq 3$~keV).
These pieces of evidence are in agreement with the
conclusions drawn by Ferrarese \& Ford (1999), who
remark that the {\it nuclear} non-thermal bolometric
optical luminosity in NGC~6251 is larger than predicted
for accretion at the Bondi rate with 10\% efficiency,
making the case for an ADAF far less compelling.

\subsection{Additional components in the nuclear X-ray spectrum
of NGC~6251}

As already noticed by Turner et al. (1997),
a soft X-ray excess is observed in the NGC~6251 ASCA
spectra.
A multitemperature blackbody is not able
to fit the excess, arguing against the possibility
that this component
originates in an accretion disk. The
best-fit temperature of the soft excess,
if modeled with an optically thin, collisionally
excited plasma, is
$kT \simeq 0.8$~keV, with an
unabsorbed luminosity of $\simeq 3.6
\times 10^{41}$~erg~s$^{-1}$. Such a temperature
is typical of gaseous halos in elliptical galaxies
(\cite{matsumoto97}).
The parameters of this component are consistent
with those of a similar component in the BeppoSAX
spectrum, which is, however, not required from
the statistical point of view. 
The ASCA temperature is in turn consistent, within the
rather large statistical uncertainties,
with the temperature of the extended halos
measured by the {\it Einstein}/IPC and
the ROSAT/PSPC in NGC~6251 (for the latter, $kT = 0.3$--0.8~keV;
\cite{birkinshaw93} and references therein).
Similarly, the 0.5--2.4~keV unabsorbed fluxes of
this component measured by ASCA and BeppoSAX
are consistent
with the ROSAT/PSPC soft X-ray "halo"
flux, integrated across its whole extension
[$(2.4 \pm ^{4.2}_{2.0}) \times 10^{-13}$~erg~cm$^{-2}$~s$^{-1}$;
\cite{mack97}].
It is therefore straightforward to identify the
soft excess in the large ASCA aperture with the
diffuse emission in ROSAT.
Mack et al. (1997) discuss (and rule out)
the possibility that such a plasma could be the
confining medium for the jet, along the
whole structure up to knot ``D". A $\simeq$$10^7$~K
thermal emission responsible for
the confinement of the jet in NGC~6251 should
not extend beyond scales larger than 60~pc.
On such small scales, it might be
marginally resolvable by {\it Chandra}.
Unfortunately, the pile-up affecting
the {\it Chandra} observation
prevented us from deriving precious constraints on
the X-ray extension around the NGC~6251 core.

The ASCA spectrum unveils the possible presence of a third
spectral component, whose main signature is a bright
($EW \simeq 600$)~eV and significantly ionized
K$_{\alpha}$ fluorescence iron line. This component,
even if present with comparable flux in the
BeppoSAX  observation, would have
easily missed detection, due to the
fact that the BeppoSAX non-thermal continuum 6~keV
flux density was $\ge$6 times larger than
in ASCA. This "third" component cannot
be produced by the superposition of unresolved
discrete sources, integrated in the large
ASCA aperture. The {\it Chandra} image
shows only two sources in the innermost 4$\arcmin$
around the NGC~6251 core, whose total count rate
[($1.1 \pm 0.3) \times 10^{-3}$], corresponds
to a 0.5--9~keV flux
$\simeq$$1.0 \times 10^{-14}$~erg~cm$^{-2}$~s$^{-1}$
(assuming a thermal plasma with $kT = 6$~keV). This is
more than one order of magnitude less than the
flux of the ASCA component in the same energy
band ($2 \times 10^{-13}$~erg~cm$^{-2}$~s$^{-1}$).
This component may therefore represent
the "tip of the iceberg" of an underlying Seyfert
nuclear "warm scattering" (Matt et al. 1996, 2001),
whose relative contribution
becomes not-negligible during phases of lower
blazar activity. Alternatively,
this component may represent a hotter and weaker
phase of the thermal
emission. It has already been pointed out that
the kpc-scale jet structure in NGC~6251 requires
an ambient gas with temperatures of 2--5~keV
to confine it. So far, the lack of detection
of a gas component with such an high-temperature
had led to the problem of an ``over-pressurized"
jet in NGC~6251. It would be tempting to
speculate that the high-temperature thermal
component measured by ASCA is
the required "missing link" to ensure the
thermodynamical stability of the kpc-scale jet.
Unfortunately, only rather strict upper limits can be
set on any X-ray emission associated with the
radio jet (see Table~\ref{tab2}).
Alternatively, the line could be produced in the
interaction between the "bloated" base of
a jet with a stratified velocity structure
and the circumnuclear matter, as proposed
by Chiaberge et al. (2000) in the context of a
possible unification model between BL Lac
objects and FR~I radio galaxies.
The possibility that the iron
line is associated with the highly
ionized ``skin" of an accretion disk
cannot be in principle ruled out, although
evidence for significantly ionized disk lines
is rather scanty so far.

All the best-fit X-ray models
discussed in this paper
require an amount of cold photoelectric
absorption in excess to the Galactic contribution.
The weighted between the
BeppoSAX and ASCA measurements
is $N_H = (1.06 \pm 0.11) \times
10^{21}$~cm$^{-2}$.
This is in excellent agreement with the column density
through the dusty disk, as derived from its visual
extinction ($A_V = 0.61 \pm 0.12$; \cite{ferrarese99}),
if standard
gas-to-dust Galactic ratios are assumed. This is in agreement
with the idea that the standard pc-scale optically
and geometrically thick torus - even if present in FRI
low-luminosity radio galaxies - does not intercept our
line-of-sight to the nucleus (\cite{chiaberge02}).

\begin{acknowledgements}

This paper benefitted of the stimulating scientific
environment at the Workshop "AGN spectroscopy with
{\it Chandra} and XMM-Newton", held at the Max Planck
Institut f\"ur Extraterrestrische
Physik in Garching. Support by Maria Teresa Fiocchi
in using the SSC models fitting facility at the A.S.I.
Science Data Center is gratefully acknowledged.
Last, but not least, comments by the
referee (Dr. M. Boettcher) greatly helped us
to sharpen the focus of the discussion.

\end{acknowledgements}

\end{document}